\documentclass[]{spie}  

\usepackage{amsmath,amsfonts,amssymb}
\usepackage{graphicx}
\usepackage[colorlinks=true, allcolors=blue]{hyperref}

\usepackage[numbers,sort&compress]{natbib}
\usepackage{aas_macros}

\title{The SHARPEx space-to-space VLBI experiment}

\author[a]{Freek Roelofs}
\author[a]{Christiaan Brinkerink}
\author[b]{Albert Catalan Artigas}
\author[a]{Heino Falcke}
\author[c]{Jaime Fernández}
\author[d,e]{Christian M. Fromm}
\author[d]{Felix Glaser}
\author[a]{Marc Klein Wolt}
\author[f]{Yuri Kovalev}
\author[g,h]{Volodymyr Kudriashov}
\author[b]{María Manzano}
\author[i]{Rasmus Maråk}
\author[g]{Manuel Martin-Neira}
\author[j]{Jacobo Pastor Fernández-Posse López}
\author[b]{Montserrat Puertolas Turell}
\author[j]{Alberto Rodríguez Pérez-Silva}
\author[k]{René Rüddenklau}
\author[l,f]{Efthalia Traianou}
\author[b]{Roger Vilaseca Miro}
\affil[a]{Department of Astrophysics, Institute for Mathematics, Astrophysics and Particle Physics (IMAPP), Radboud University, P.O. Box 9010, 6500 GL Nijmegen, The Netherlands}
\affil[b]{SENER AEROESPACIAL, Carrer Creu Casas i Sicart, 86-88, Parc de l'Alba, 08290 Cerdanyola del Vallès, Barcelona (España)}
\affil[c]{GMV AD., Tres Cantos, Spain}
\affil[d]{Institute for Physics and Astronomy, University of Würzburg, Emil-Hilb-Weg 31, 97074 Würzburg, Germany}
\affil[e]{Institute for Theoretical Physics, Goethe Universität Frankfurt, Max-von-Laue-Str. 1, 60438 Frankfurt, Germany}
\affil[f]{Max-Planck-Institut für Radioastronomie, Auf dem Hügel 69, D-53121 Bonn, Germany}
\affil[g]{European Space Agency, Keplerlaan 1, 2200-AG Noordwijk, The Netherlands}
\affil[h]{Serco Netherlands BV}
\affil[i]{Scaleout Systems, Kungsgatan 12, 753 32 Uppsala, Sweden}
\affil[j]{Alén Space, Edificio Tecnológico Aeroespacial, Rúa das Pontes, 6, 2.05, 36350 Nigrán, Pontevedra, Spain}
\affil[k]{Institute of Communications and Navigation, German Aerospace Center (DLR), Münchenerstr. 20, 82234 Weßling, Germany}
\affil[l]{Instituto de Astrofísica de Andalucía-CSIC, Glorieta de la Astronomía s/n, E-18008 Granada, Spain}

\authorinfo{Further author information: (Send correspondence to F.R.)\\F.R.: E-mail: f.roelofs@astro.ru.nl, Telephone: +31 24 365 2804}

\begin{document} 
\maketitle

\begin{abstract}
Space is the next frontier for ultra-high resolution imaging of astrophysical sources with very long baseline interferometry (VLBI). The limited size of the Earth and atmospheric absorption and turbulence at high frequencies prevent order-of-magnitude resolution improvements from current black hole imaging with the ground-based Event Horizon Telescope. A two or three-element submm space-based array like SHARP would image the thin black hole photon ring and measure the black hole spin and spacetime. Here, we present the SHARP Experiment (SHARPEx) mission concept. SHARPEx will demonstrate key enabling technologies for space-to-space VLBI, including clock syntonization between two satellites, on-board correlation, and sub-wavelength relative orbit determination. SHARPEx will consist of two CubeSat or SmallSat platforms carrying small ($\sim$1 m diameter)  antennas, observing at cm wavelengths. While SHARPEx will be a technical demonstrator, a SHARPEx-type follow-up mission with enhanced intersatellite link properties (range and bandwidth) may produce images of bright AGN with unprecedented resolution and fidelity.
\end{abstract}


\section{INTRODUCTION}
\label{sec:intro}
\subsection{From ground to space-based VLBI}
The astronomical observing technique of very long baseline interferometry (VLBI) at radio wavelengths has been highly successful, achieving the highest resolutions of any imaging experiment. The Event Horizon Telescope (EHT), a ground-based global VLBI array operating at $\sim$1 mm wavelengths, has made the first images of black holes at a resolution of 20 $\mu$as \citep[e.g.,][]{M87PaperI, SgrAEHTCI}. 

\begin{figure}[t]
    \centering
    \includegraphics[height=0.35\textwidth]{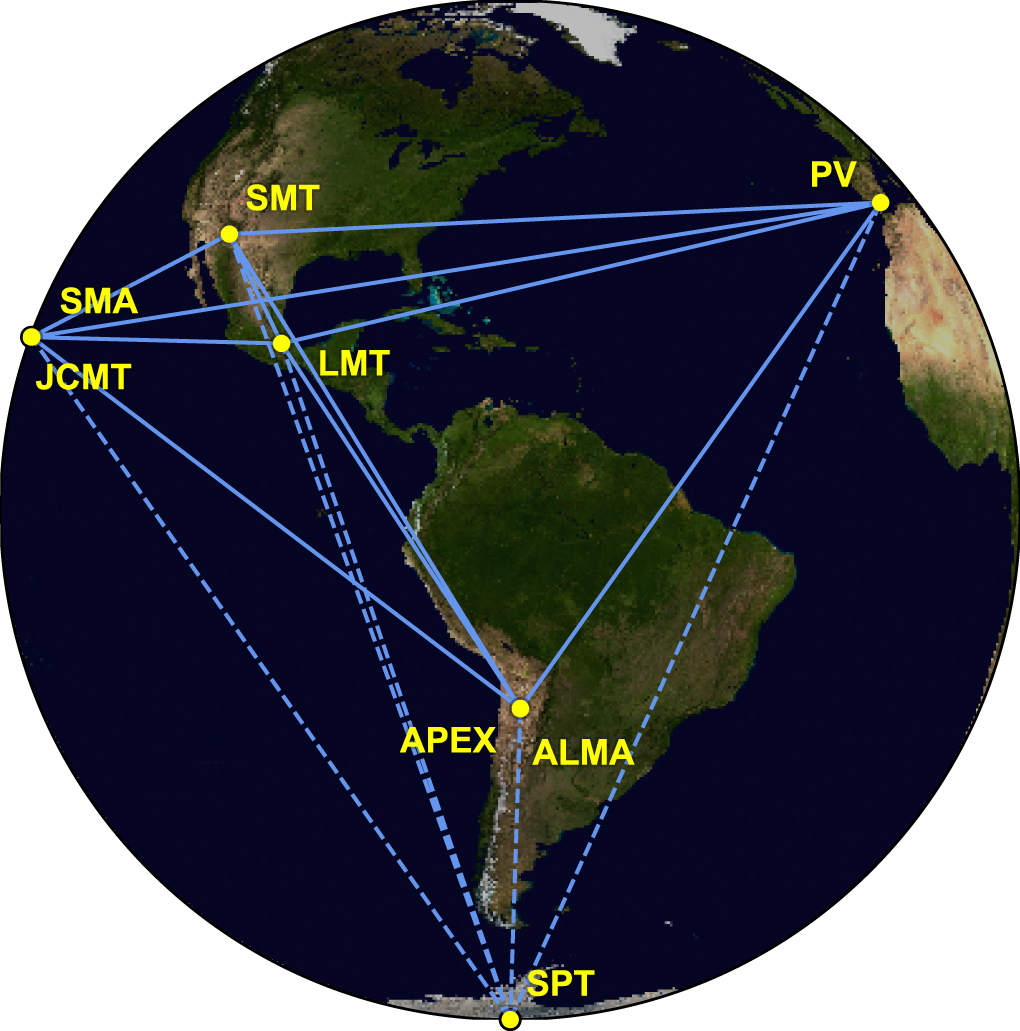}
    \includegraphics[height=0.35\textwidth]{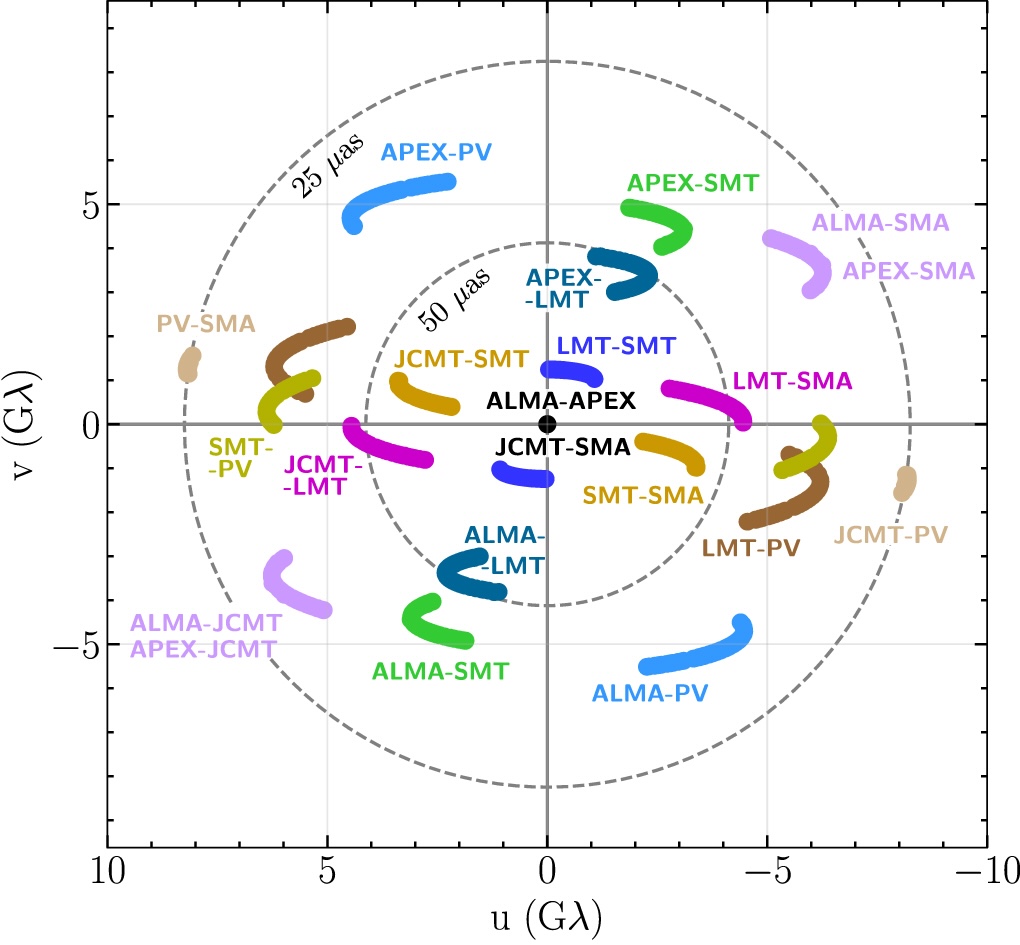}
    \includegraphics[height=0.35\textwidth]{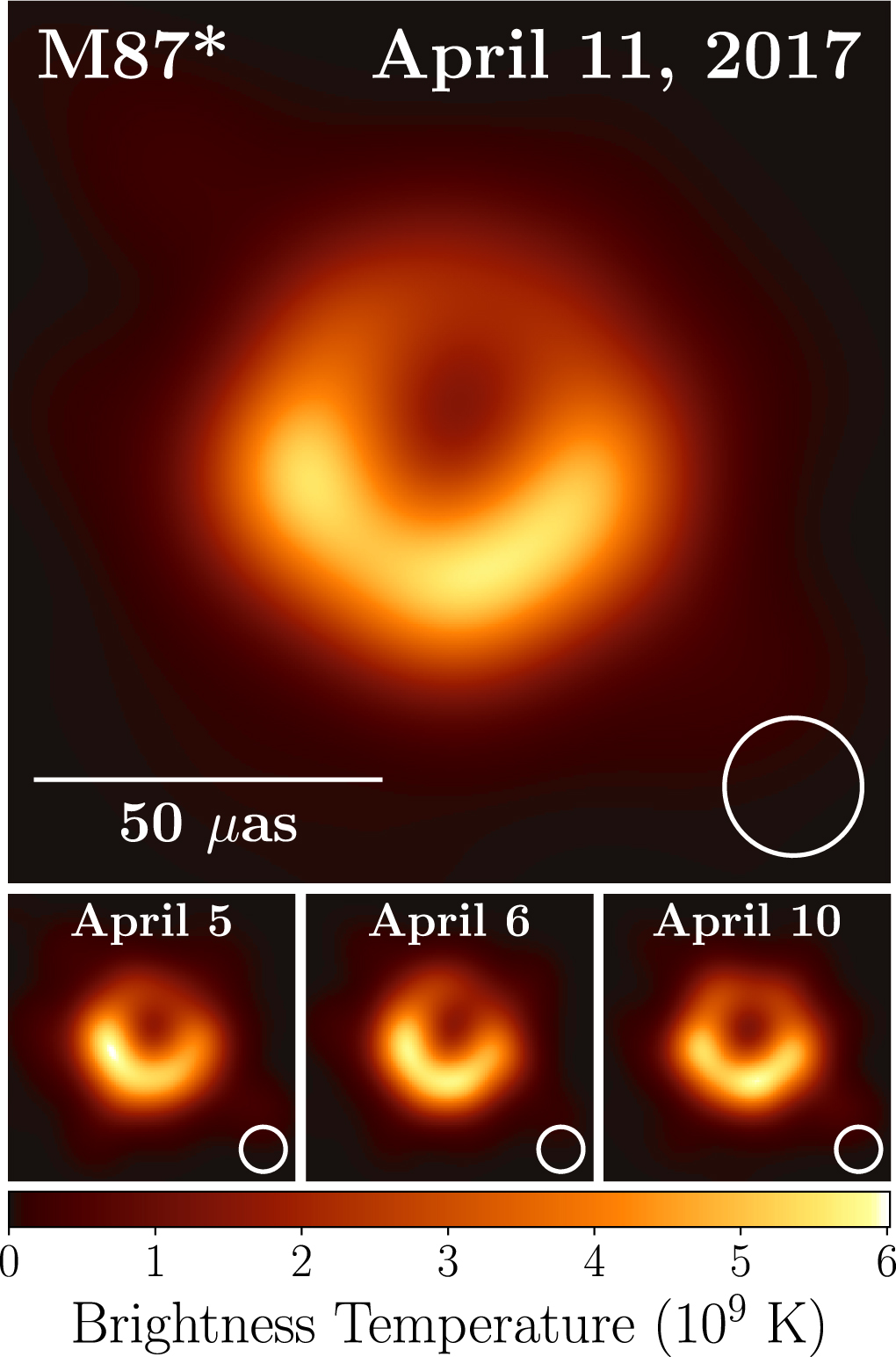}
    \caption{VLBI imaging from the ground: baseline lengths are limited by the size of the Earth and attainable frequency (the atmosphere severely absorbs emission at $\gtrsim 230-345$ GHz), and gaps in the $uv$-coverage are inevitable due to the limited number of telescopes. Left: the EHT array in 2017. Middle: EHT $uv$-coverage of M87* in 2017. Right: first EHT images of M87 at 230 GHz. Panels reproduced from \citep{M87PaperI}.}
    \label{fig:eht2017}
\end{figure}

A VLBI array like the EHT samples Fourier components (``visibilities'') of the observed image on a limited set of baselines (projected distances between two telescopes). The resolution of the array is set by the longest baseline measured in observing wavelengths. The $uv$-coverage, or distribution of the sampled baseline lengths and directions in a 2D-plane, determines image fidelity: sparse $uv$-coverage with many gaps (due to, e.g., a small number of telescopes in the array) will not give sufficient Fourier components to reconstruct a reliable image. In Earth-rotation synthesis as performed by the EHT and other VLBI arrays, the projected baselines sweep out ellipses in the $uv$-plane as the Earth rotates during an observation, filling the $uv$-plane sufficiently for reliable image reconstruction (\autoref{fig:eht2017}).

While the dynamic range and image fidelity of the EHT can be improved by adding more telescopes \citep[e.g.,][]{Backes2016, Roelofs2020, LaBella2023, Roelofs2023, Doeleman2023, Johnson2023}, astronomical imaging beyond the resolution of the EHT will be extremely challenging due to the limited size of the Earth and atmospheric absorption and turbulence, which become increasingly severe towards shorter wavelengths. Both these limitations can be overcome with a space-based VLBI array, which allows for larger distances between telescopes and higher-frequency observations due to the absence of an atmosphere. By picking suitable orbits, the $uv$-plane can be filled completely even with only two satellites, resulting in excellent imaging capabilities (see the SHARP concept and \autoref{fig:uvcov} below, obtaining a 3.5 $\mu$as resolution).

\subsection{Space-to-space VLBI science goals: black hole and jet imaging}
A space-based array with ultra-high resolution and fidelity imaging capabilities will be extremely valuable for astrophysics and fundamental physics. For the supermassive black holes M87* and Sgr~A*, extreme light bending around a black hole into a thin ``$n=1$ photon ring’’ will be visible at resolutions beyond $\sim$10 $\mu$as \citep[e.g.][]{Johnson2020}. This thin ring consists of light emitted near the black hole that has made a U-turn around the black hole before reaching the distant observer. Towards higher frequencies approaching the THz regime, the black hole image will be increasingly dominated by this photon ring, which approaches the boundary around the dark black hole ``shadow'' \citep{Falcke2000}. Unlike the direct emission, which dominates current EHT images and whose appearance is strongly determined by the detailed properties of the accreting plasma, the size and shape of the thin $n=1$ ring are predominantly set by the spacetime metric of the black hole itself. Measuring this photon ring will thus provide unprecedented direct constraints on the black hole spin, and allow for precise tests of theories of gravity in an unexplored regime. Furthermore, increasing the image resolution will allow imaging the shadows of more black holes, depending on the sensitivity of the space array. Several black hole shadows exists with a diameter between 5 and 10 $\mu$as and flux densities between 10 and 100 mJy at 230 GHz \citep{Ramakrishnan2023}. Potential targets are M84, M104, 3C270, and NGC3998.

Black holes accrete matter, which heats up and forms a strongly magnetized plasma. Near (supermassive) black holes, the magnetic field is advected onto and wound up near the horizon, causing the formation of large-scale jets which launch particles far into the intergalactic medium. These extremely high-energetic jets are thought to be powered by the spin of the black hole itself, in a spin energy extraction mechanism called the Blandford-Znajek process \citep{Blandford1977}. With polarimetric images and movies of black holes at a resolution of a few $\mu$as, this jet launching process can be seen directly, linking spacetime to plasma physics.

To complete our picture of jet launching and understand plasma behavior, we need sharp images of jets on larger scales as well. Detailed images showing edge brightening, instabilities, and inner spine emission will increase our understanding of the entire system. Contrary to the photon ring images, the large-scale jet emission shows up at lower frequencies (cm wavelengths). A pathfinder for or lower-frequency version of a high-frequency space VLBI mission has the potential to image jets in unprecedented detail that is unattainable from the ground (see \autoref{sec:sims}).

\begin{figure}[t]
    \centering
    \includegraphics[width=0.9\textwidth]{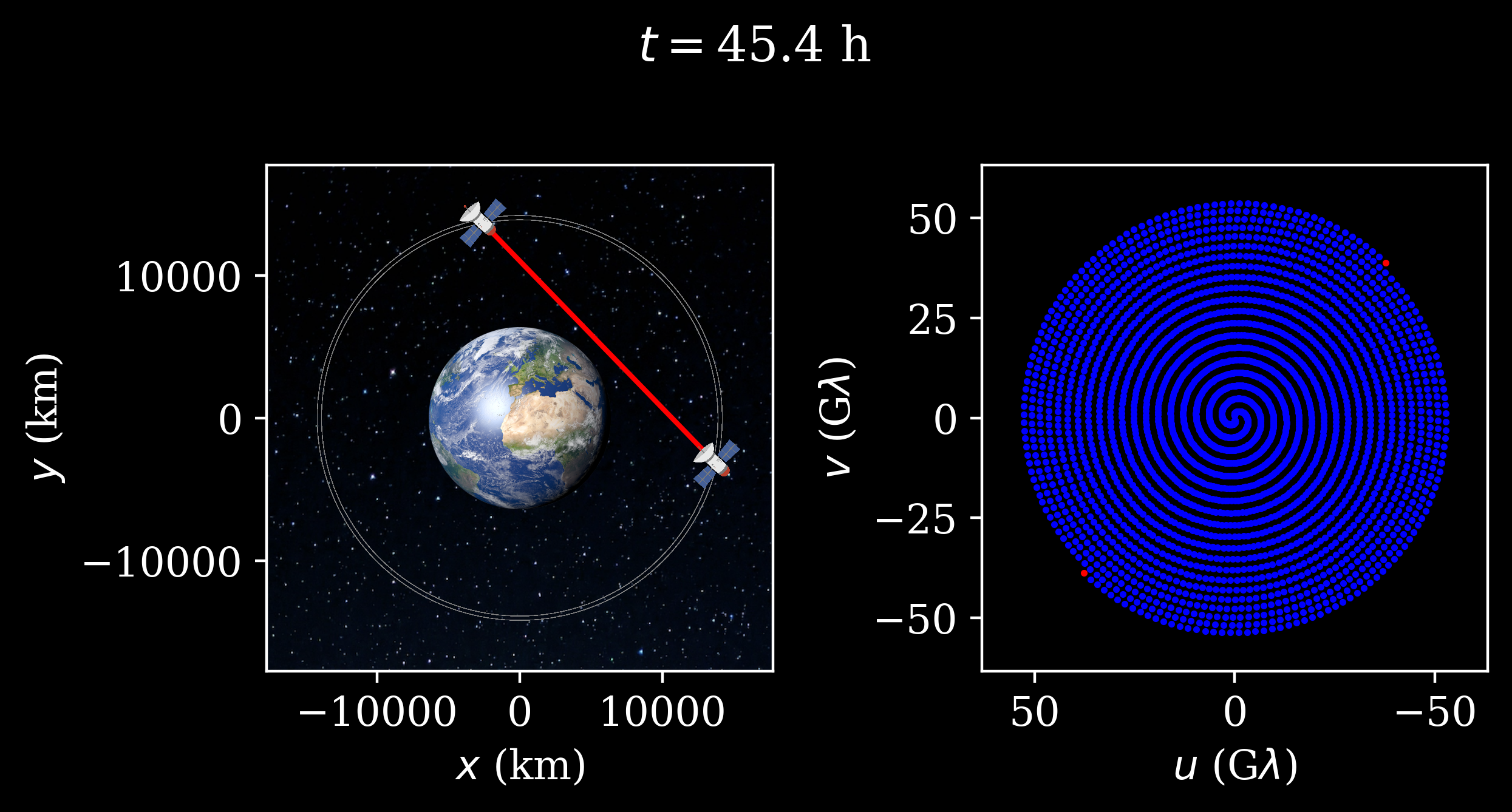}
    \caption{Video 1. Illustration of SHARP orbits and $uv$-coverage at 690 GHz in case of a two-element array \citep[see also][]{Roelofs2019}. The relative drift of the orbiting satellites results in a dense spiral-shaped $uv$-coverage allowing for excellent imaging up to a resolution of 3.5 $\mu$as. Earth image by \href{https://www.solarsystemscope.com/spacepedia/handbook/earth}{Solar System Scope}. Animation available at http://dx.doi.org/10.1117/12.3104732.1.}
    \label{fig:uvcov}
\end{figure}

\subsection{The SHARP space-to-space VLBI concept}
\begin{figure}[t]
    \centering
    \includegraphics[width=0.45\textwidth]{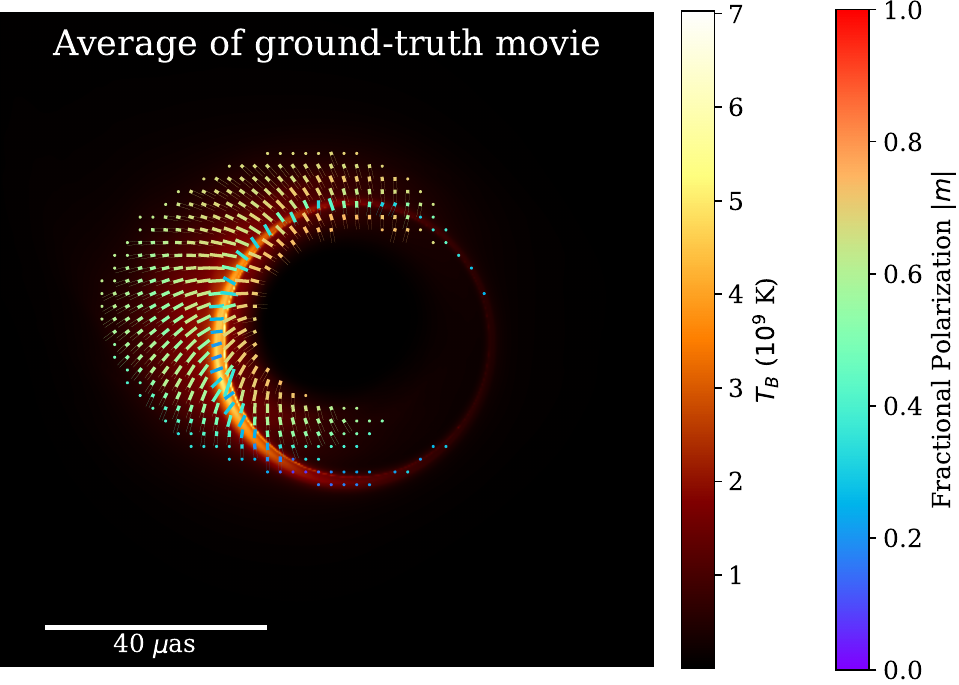}
    \includegraphics[width=0.45\textwidth]{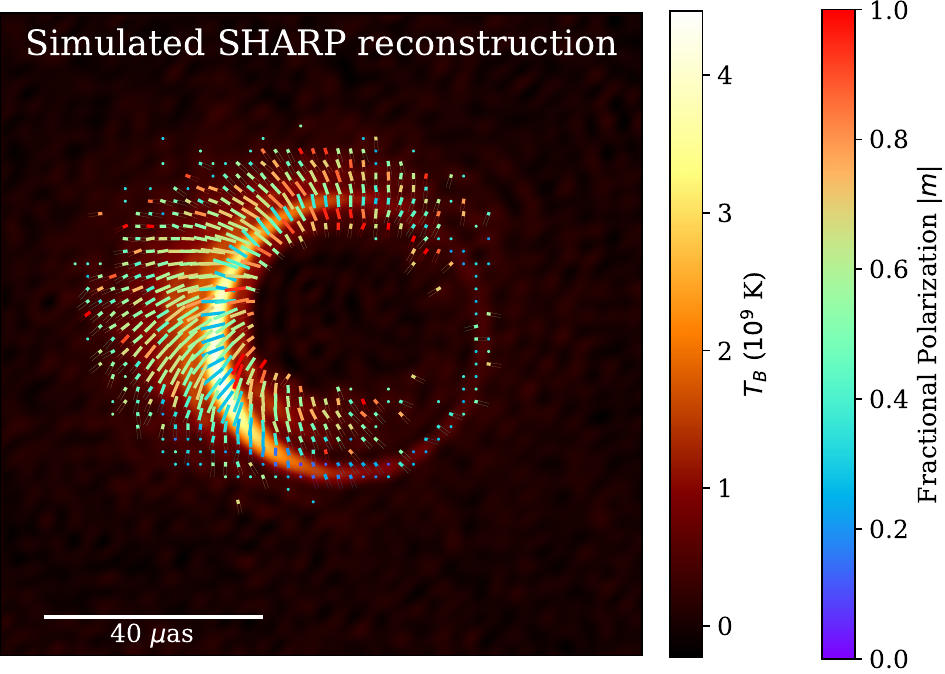}
    \caption{Left: time-averaged image of the GRMHD movie (MAD, $a_*=0.9375$, $R_{\mathrm{low}}=1$, $R_{\mathrm{low}}=160$) used as input for synthetic SHARP observations of Sgr~A* \citep{Glaser2026}. Right: image reconstruction (FFT) from simulated SHARP data, recovering the thin $n=1$ photon ring in both total intensity and polarization. The reduced fractional polarization on the $n=1$ ring is a consequence of the fact that light makes a U-turn around the black hole, leading to a flip in the EVPA pattern handedness and depolarization when combined with the direct emission, which has opposite EVPA pattern handedness \citep[see also][]{Jimenez2021,PalumboWong2022}. The synthetic observations were performed using the \texttt{svlbisim} tool\protect\footnotemark with three satellites carrying 3.4 m diameter dishes, with a system temperature of 150 K, bandwidth of 10 GHz, 3-bit sampling, and aperture efficiency of 0.7, observing for a total of six months.}
     \label{fig:sharpsims}
\end{figure}

In order to achieve the science goals described above, we have developed the SHARP mission concept and ESA M-class mission proposal, inheriting critical ideas from the Event Horizon Imager (EHI) space-to-space VLBI concept \citep{Martin2017, Roelofs2019, Kudriashov2021}. SHARP will consist of two or three satellites in Medium Earth Orbits (MEOs), attaining resolutions of 3.5 $\mu$as when observing at 690 GHz (the actual observing frequency of SHARP is to be determined in a detailed science requirements study, but will likely be in the 500+ GHz regime). Due to slight differences in orbit radii and consequential drifts, baselines will be sampled from low to high resolutions and in all directions, allowing extremely high-fidelity imaging (\autoref{fig:uvcov}). The maximum baseline length is set by the point where the Earth occludes the line of sight between two satellites, since the data and clock will need to be shared over an intersatellite laser link (see \autoref{sec:challenges}). Simulations have shown that such a concept will produce extremely high-resolution and high-fidelity images of black holes, allow for the measurement of black hole spin, and even produce high-definition movies of black hole accretion processes \citep[][and \autoref{fig:sharpsims}]{Roelofs2019, Roelofs2021, Shlentsova2024}. 

Besides SHARP, other Space VLBI concepts to achieve similar or related science goals are under development. The Black Hole Explorer (BHEX) concept entails a single-satellite extension of the ground-based EHT, operating at frequencies up to 320 GHz, to detect and measure a black hole's photon ring \citep{Johnson2024}. Like SHARP, Capella \citep{Trippe2023} is a space-to-space concept operating at $\sim$690 GHz, but consisting of four antennas in Low Earth Orbits. Beyond SHARP, a larger constellation or Space Array concept would open the door to high-sensitivity observations with fast $uv$ filling, with a broader set of science applications (e.g., protoplanetary disks, transients).

\section{Key technical challenges and solutions}
\label{sec:challenges}
In the SHARP concept, raw high-bandwidth ($\sim$tens of Gbps) data is shared between the satellites through an intersatellite laser link (ISL). The data is then correlated on-board before being sent to Earth for further processing. In order to perform the correlation with a narrow delay and delay-rate window and hence achieve a significant data reduction before downlinking the data, the relative 3D orbits (position, velocity, acceleration) must be known to high precision. GNSS satellites may provide cm-level positions in real time, which would reduce the data sufficiently to utilize RF downlinks \citep{Kudriashov2021}, so that the concept will not need to depend on a Gbps capacity optical downlink and a ground-based network of optical terminals.

On long baselines, the visibility amplitudes of the primary science targets of SHARP are expected to be too low for a conventional VLBI fringe detection on timescales of minutes, since the concept is restricted to small dishes and hence the thermal noise will be significantly larger than for the large ground-based dishes typically used in VLBI. Fringe detection requirements must therefore be strongly mitigated, so that low signal-to-noise data may be fully utilized to obtain information on the source structure and signal-to-noise can be built up over longer timescales. In the SHARP concept (\autoref{fig:uvcov}), the same approximate ($u,v$) points are revisited multiple times as the satellites drift apart and catch up again on a timescale of days.

\begin{figure}[t]
    \centering
    \includegraphics[width=0.7\textwidth]{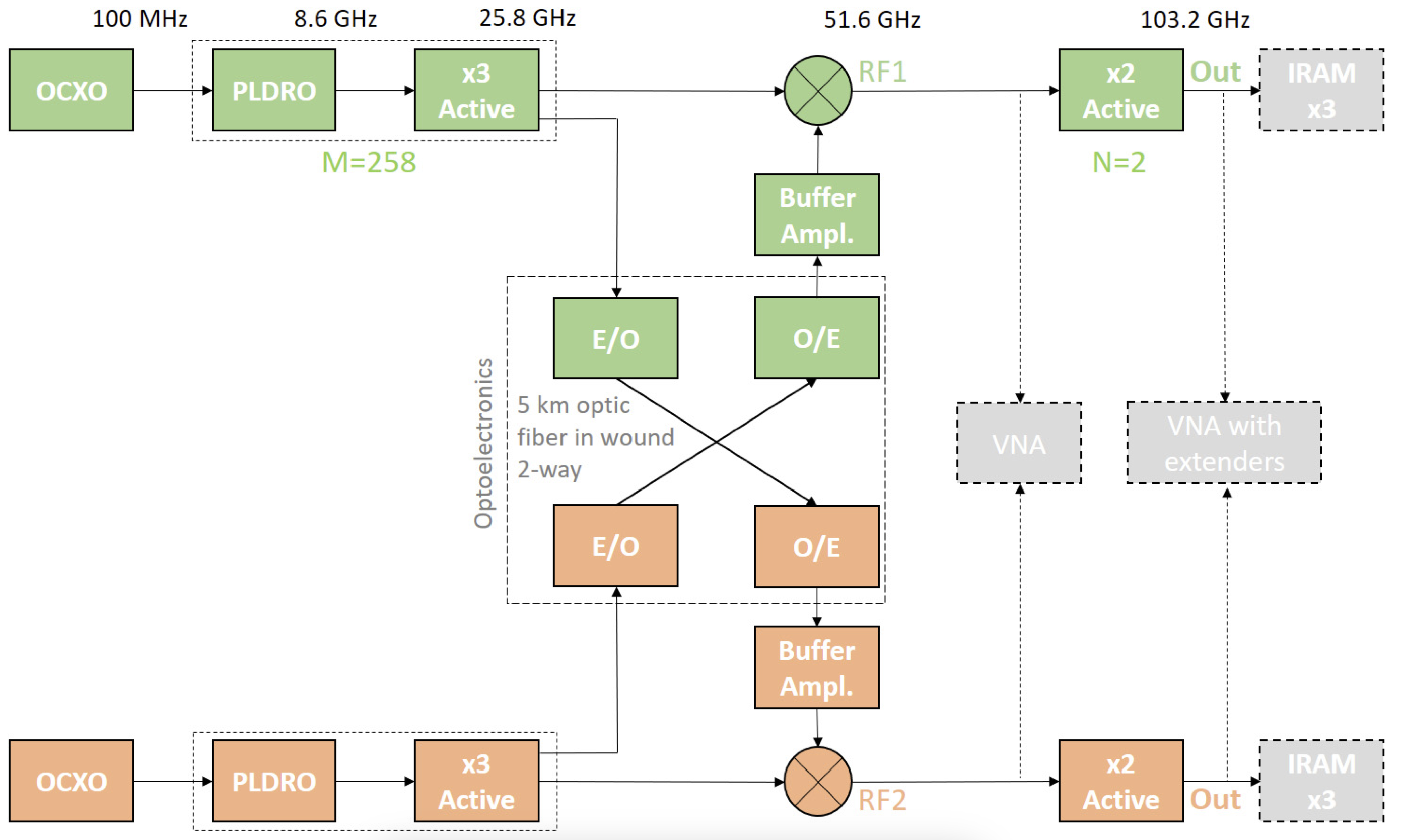}\\
    \includegraphics[width=0.35\textwidth]{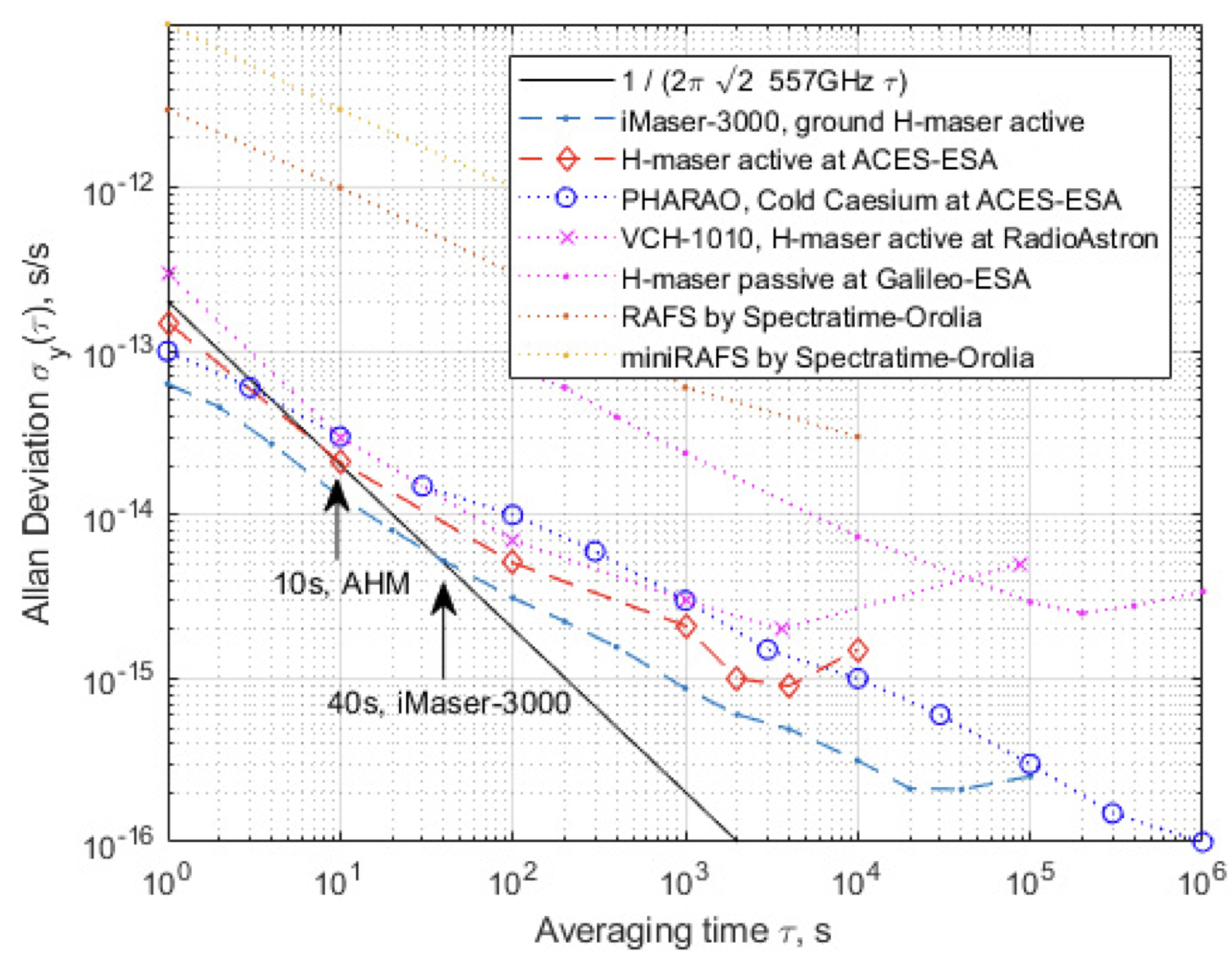}
    \includegraphics[width=0.35\textwidth]{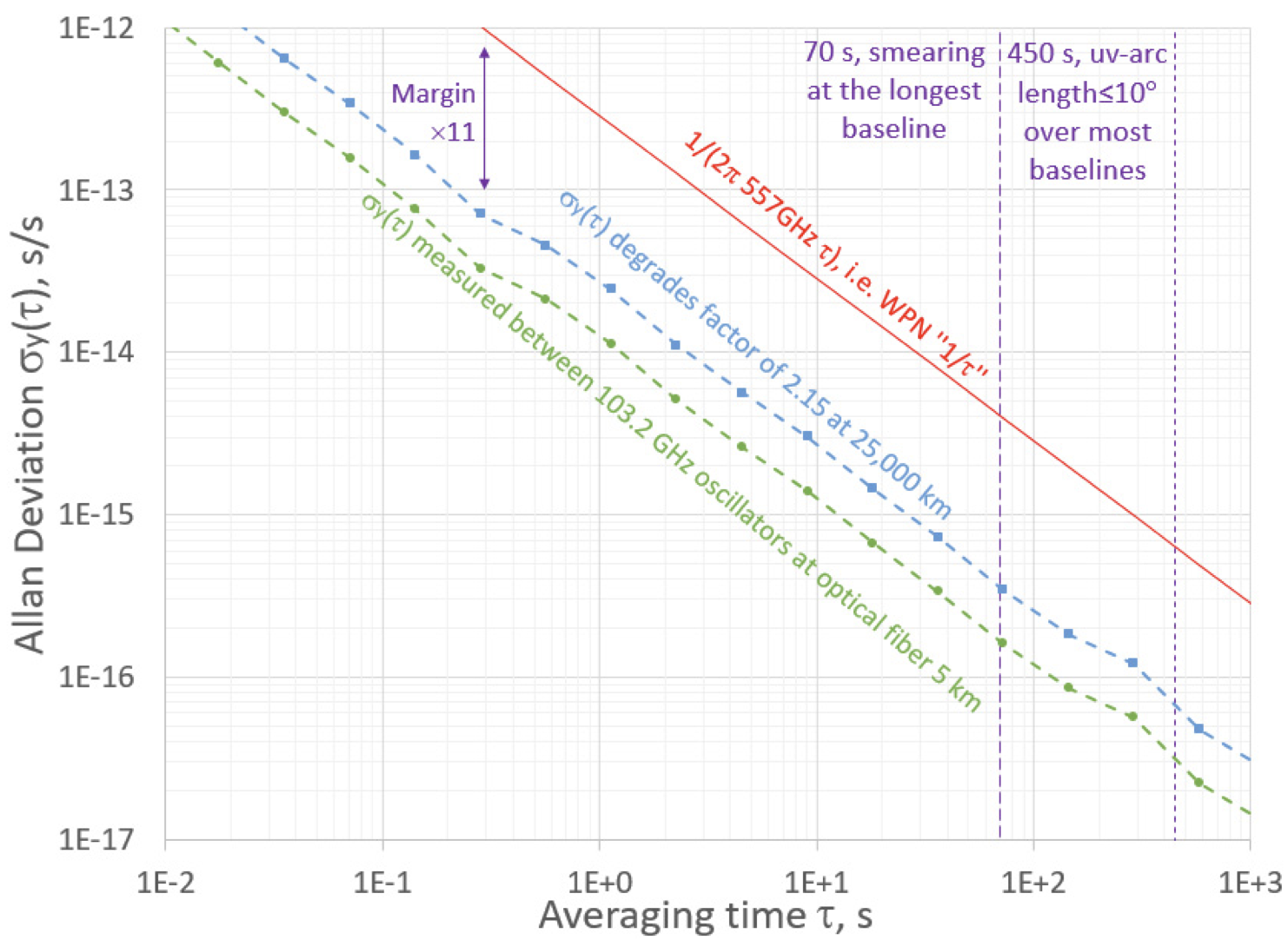}
    \caption{Top: high-level block diagram of the clock syntonization system demonstrated in the lab at ESTEC. Bottom left: Allan Deviation for several masers, indicating a coherence time of at most tens of seconds at 500+ GHz. Bottom right: Allan Deviation measurements using the clock syntonization system, demonstrating excellent coherence to at least 1000 seconds with no signs of significant degradation towards longer integrations. Panels reproduced from \citep{Kudriashov2021lo}.}
     \label{fig:clocks}
\end{figure}

Building up signal-to-noise over long timescales requires excellent interferometric phase coherence, much like a connected interferometer on the ground. In the absence of an atmosphere, clock errors would pose the main limitation on the phase coherence time, as space-qualified masers typically lose the required coherence after tens of seconds. The SHARP concept solves this issue by creating a highly phase-stable interferometer by exchanging the clock signals between two satellites. Mixing the local with the incoming clock signal at each satellite, a fully syntonized clock is created, allowing long integrations \citep[][\autoref{fig:clocks}]{Kudriashov2021lo}. 

Apart from clock phase stability, the other main requirement for a phase coherent interferometer is sub-wavelength level knowledge of the 3D relative vector between the satellites. Such precision may be reached by post-processing of the GNSS observables used for the real-time correlation, potentially aided by measurements from accelerometers. ISL ranging measurements help constraining the solution in one direction. From simulations, a 3D baseline measurement with an error of 3 mm after postprocessing seems possible for SHARP in MEO about 60\% of the time with postprocessing of GNSS data alone \citep{Moradi2022, Salas2024}\footnote{ESA Activity NAVISP-EL1-024}.

While the attainable precision of combining different relative navigation techniques (GNSS + accelerometers) remains to be studied in more detail, SHARP may also need to rely on simultaneous multi-band observations. The phase solutions obtained from detecting the source at a lower frequency band, where the signal-to-noise ratio is typically higher and relative navigation requirements are less stringent, may be propagated to the higher science frequency band using the Frequency Phase Transfer technique \citep[FPT, e.g.][]{Middelberg2005, Rioja2011}. The applicability of this technique to the SHARP science targets will also be investigated further. 

\footnotetext{\url{https://github.com/freekroelofs/svlbisim}}

\section{THE SHARPEx DEMONSTRATOR CONCEPT}
\label{sec:objectives}
The space-to-space VLBI concept described above is a completely new way of observing the universe, and comes with challenging system requirements. These requirements can be met with key enabling technologies that will need to reach sufficient Technology Readiness Levels (TRLs) before a mission like SHARP can fly. Here, we therefore propose the SHARP Experiment (SHARPEx). SHARPEx will be a demonstrator of critical enabling technologies for SHARP. SHARPEx will demonstrate:
\begin{itemize}
    \item The technical feasibility of performing VLBI between two antennas in space.
    \item On-board correlation on a space-to-space interferometric baseline.
    \item Relative navigation solutions for sub-wavelength orbit determination.
    \item Clock syntonization and synchronization between two satellites.

\end{itemize}

\subsection{Envisioned SHARPEx concept and study}
\noindent SHARPEx will observe bright AGN at cm wavelengths. Observing at lower frequencies mitigates requirements on the relative navigation (orbit reconstruction), clock stability, and sensitivity compared to the full-scale SHARP concept, while still demonstrating its basic principles in-orbit. We envision SHARPEx to consist of two CubeSats carrying deployable antennas with a diameter of about 1 m. The raw data will be shared over an intersatellite link and correlating the data on-board before sending it to the ground, like in the SHARP concept. We are starting a systems study under the ESA OSIP Programme\footnote{ID I-2025-05960; \url{https://ideas.esa.int/core/servlet/hype/IMT?userAction=Browse&templateName=&documentId=45583922facac645dac74c7650ca3f0e}} to investigate the feasibility of such a configuration for our demonstration goals, and to narrow down its system parameters. The outcome of this study will form the basis for a mission proposal.

\subsection{Minimum objectives}
\noindent To demonstrate the SHARP space VLBI concept, SHARPEx will need to observe and process interferometric data with at least the following properties:
\begin{itemize}
    \item The observed source needs to be partly resolved. This means that the correlated flux density (visibility amplitude) must be shown to decrease as a function of baseline length.
    \item The visibility phase needs to be calibrated, and shown to depend on the baseline length and/or orientation as a condition for resolving the source.
    \item When observing a non-variable source, the phase needs to be the same when returning to the same baseline for multiple iterations of the uv-spiral. Such phase stability implies that the 3D baseline ($u, v, w$) must be known to within an observing wavelength in postprocessing, and that the clocks on both satellites need to be syntonized.
\end{itemize}

A mission like SHARPEx has the potential to lead to a high-impact scientific mission at cm wavelengths as well. However, a science mission producing excellent astronomical images beyond what is possible from the ground would require baseline lengths exceeding an Earth diameter and, for $\sim 1$ m antennas, an observing bandwidth of $\sim$ a few hundred MHz, bringing such a mission outside the CubeSat envelope (see \autoref{sec:sims}).

\subsection{System requirements and critical technology developments}
\noindent We describe the envisioned system components and open questions to be investigated below. Not all these components and issues will be fully addressed in our OSIP study, but we highlight what we consider to be the most important aspects.

\begin{itemize}

    \item \underline{Intersatellite link (ISL):} The measurement of correlated flux density requires the combination of sampled radio data from the two spacecraft platforms in order to perform cross-correlation on the two data streams. This correlation can either be done on Earth by having each of the spacecraft first send its sampled data down, or alternatively by having one of the spacecraft send its sampled data to the other platform over an intersatellite link for on-board correlation. This latter option has our preference (see the point on on-board correlation below). 
    
    The ISL bandwidth and distance necessary for reliable detection and characterization of astrophysical sources by SHARPEx is a parameter to be converged upon. The imaging performance of SHARPEx will strongly depend on these parameters (see \autoref{sec:sims} and \autoref{fig:ehipimaging} for a simulation test case where various ISL bandwidths and distances were used). Making data exchange across baseline lengths of thousands of kilometers possible with high bandwidths likely requires laser ISLs, but for SHARPEx RF links are also considered. An example laser ISL developed for CubeSats is CubeISL, which can establish a 0.1 Gbit/s data rate between CubeSats separated by a distance of 1500 km \citep{rodiger2025cubeisliod}. Its successor Cube1G \citep{Rodeck2025} adds a Coarse Pointing Assembly, so that the ISL terminal is able to point to the other satellite independently of the orientation of the main antenna and GNSS receiver.

    In addition to the wide-band interferometric data, the ISL will also be used to share the clock signals (see clock syntonization below), as well as telemetry, telecommand, and relative navigation information. The ISL can also be used for ranging measurements, although previous studies have indicated that the added improvement of the 3D relative positioning by ISL ranging measurements may be marginal \citep{Salas2024}.
    
    \item \underline{Precise orbit determination:} To ensure phase stability for the correlated signal, the relative 3D positions of the two spacecraft with respect to one another need to be known accurately enough for the correlator model to account for the path length changes with time, allowing for correlation with narrow delay and delay-rate windows. Such precision is attainable in real time with GNSS measurements (see also \autoref{sec:challenges}). 
    
    After postprocessing on the ground, the relative navigation needs to reach sub-wavelength precision. The observing frequency of SHARPEx is also a parameter to be converged upon. At 22 GHz, the wavelength is approximately 1.4 cm. At L-band (15-30 cm), requirements are an order of magnitude less stringent. Usage of GNSS and possibly intersatellite ranging is expected to bring the accuracy to the required level. Simulations indicate a 3D baseline measurement error of 3 mm is attainable in MEO about 60\% of the time \citep[\autoref{sec:challenges},][]{Moradi2022, Salas2024}. 
    
    We will investigate the performance of this relative navigation concept for the SHARPEx CubeSat-type platforms and for LEO, where simultaneous GNSS visibility is better than in MEO, but atmospheric drag causes additional disturbances of the baseline vector.

    \item \underline{Clock syntonization:} Maintaining coherence of the correlated signal also requires precise sampling clock syntonization between the two platforms. To this end, a clock-sharing architecture has been developed\footnote{ESA Patent 789, Syntonization of Signals Between Satellites} and has already demonstrated the required coherence in the lab \citep[see \autoref{fig:clocks} and][]{Kudriashov2021lo}. The same architecture also provides synchronization of the sampling and on-board correlation of the observation signals. The required performance will be demonstrated in-orbit by SHARPEx.

    \item \underline{On-board correlation:} Since on-board correlation  is critical for the SHARP concept to reach a low data rate to the ground, demonstrating this functionality is part of the intended scope for SHARPEx as well. The SHARPEx bandwidth will likely not pose fundamental issues for currently available on-board data processing hardware, but the intersatellite link needs to be able to support this data rate for all baseline distances. The on-board correlation scheme can in principle also be split between the platforms, where the correlation workload is split equally between the two spacecraft.

    \item \underline{Data processing platform and architecture:} The power requirements of the computing hardware for performing the on-board correlation needs to be compatible the available resources on the CubeSat platform. It is likely that a suitable FPGA-based architecture is available. Doppler correction by sample insertion, electronic delay correction, and application of the correlator model using relative positioning solutions all need to be implemented for successful correlation of the data streams.

    \item \underline{Data downlink:} After on-board correlation, the data bandwidth will be drastically reduced (a rough estimate for this data rate is $<$1 Mbit/s). This data can be transferred down to Earth with an RF link. In principle, the ISL laser terminal could also downlink (correlated or raw) data, so that a space-ground observing mode is a possibility. 
        
    \item \underline{Thermal management:} Depending on the orbit chosen for the SHARPEx satellites, different options for managing the thermal state of the receivers need to be assessed. Temperature stability is important for the phase stability of the shared clock and the receivers.  

    \item \underline{Antenna:} A deployable mesh Ka-band antenna has been developed by Oxford Space Systems\footnote{\url{https://satsearch.co/products/oss-hinged-rib-antenna-for-ka-band-communications-and-data-relay}}, with a diameter between 0.6 and 1.6 m, that can be stowed in about 2U. Such an antenna will have a sufficiently large collecting area for the demonstration purposes of SHARPEx (see \autoref{sec:sims}).
    
    \item \underline{Receivers:} A sensitive but non-cryogenic receiver suitable for deployment on a CubeSat platform will be identified. There is possible heritage from the MIRAS receivers on the SMOS mission, which operate at L-band \footnote{\url{https://earth.esa.int/eogateway/instruments/miras}}. Higher frequencies in general are not limited by TRL, but present more challenging requirements on relative navigation, clock synchronization, receiver sensitivity and data bandwidth. Nevertheless, higher frequencies do bring us closer to the intended envelope of SHARP and improve antenna directivity. While K-band (around 22 GHz) may be an achievable frequency band, more specific performance figures derived for the subsystems will inform the final choice of this parameter. 
    
    While likely not the case for SHARPEx, SHARP may need to rely on frequency phase transfer (\autoref{sec:challenges}) to find coherent phase solutions at the highest frequencies and longest baselines. This calibration technique requires simultaneous multi-frequency receivers. The feasibility of demonstrating FPT with SHARPEx (i.e., the feasibility of developing and installing simultaneous multi-frequency receivers, which are larger than single-frequency receivers, on the SHARPEx platforms) will be investigated. 

    While FPT in ground-based VLBI is usually performed after correlation and fringe fitting at the lower frequency, another idea to explore is to perform fringe fitting at the lower frequency in real time using a coarse orbit model, and update the correlator model for both the lower and higher frequencies with the obtained solutions to move the high frequency signal into a narrow search window. If the target is too weak for a short-timescale detection at the lower frequency, one could consider introducing two beams, one observing a target and another one observing a nearby calibrator, similar to VERA \citep{Kawaguchi2000}. 
    
    \item \underline{Orbits:} Subject to the performance of the ISL and the system sensitivity, orbits will be formulated that yield good uv-coverage for the desired sources to be studied, as well as a suitable range of baseline lengths. These orbits also need to be reachable given the launch opportunities available for CubeSat missions on rideshares, be in a benign radiation environment, allow a low-power downlink, and allow to build on small-platform heritage. While SHARP will operate in MEO, these practical considerations favor LEO for the SHARPEx demonstrator. A potential complication is the atmospheric drag in LEO impacting the relative navigation performance, which will be further investigated in our OSIP study.
    
    \item \underline{Platforms:} We aim to keep SHARPEx as small and cost-effective as possible. Each of the system components described above could in principle fit within 1U or, in case of the antenna and ISL terminal, a few U. The components are also not expected to require a prohibitively large amount of power. System integration is a key part of our OSIP study. Since the orbits of the two spacecraft likely needs to be controlled (after reaching the maximum baseline, the delta-height has to be reverted), the use of propulsion will be evaluated. If it turns out all components cannot be made to fit on CubeSat platforms, SmallSat platforms are an alternative option.  
\end{itemize}

\section{Performance simulations}
\label{sec:sims}
While demonstrating the space-to-space VLBI technique is the primary goal of SHARPEx, we are also interested in exploring the imaging capabilities of a SHARPEx-type array. The image quality to be obtained with the SHARPEx concept will depend on its sensitivity and its maximum baseline length, which are primarily set by the capabilities (bandwidth and range) of the intersatellite link, and the collecting area of the antenna. 

A mission concept providing scientific value in itself through excellent imaging at cm wavelengths would be able to count on even stronger support from the radio astronomical community. However, as shown below, a push towards unprecedented image quality in this frequency regime will likely bring SHARPEx well outside a CubeSat envelope. Such a push may significantly increase the time towards launch or cost for SHARPEx. However, if the science return is deemed significant by the community, a separate cm science ``SHARPEx+'' mission (ESA F or mini-F class) might be considered after SHARPEx.

As a benchmark for unprecedented image quality and resolution at 22 GHz, we consider the flown space-to-ground VLBI mission RadioAstron \citep{Kardashev2013}. This 10-meter antenna observed at frequencies up to 22 GHz, forming an array with ground-based antennas. Among other sources, it has made high-resolution images of the bright quasar 3C279, with detections on baseline lengths exceeding 100,000 km in a single direction \citep{Fuentes2023}. While such long baseline lengths are considered out of scope for SHARPEx or even SHARP, our imaging simulations described below show that the uniquely dense and isotropic uv-coverage of the SHARPEx concept allow obtaining an image quality exceeding that of RadioAstron with substantially shorter baselines and smaller dishes, if the ISL capabilities are significantly enhanced towards a ``SHARPEx+'' concept. In addition, the SHARPEx concept is fully independent of ground-based scheduling and weather constraints.

As a starting point for our imaging simulations, we take the parameters of CubeISL/Cube1G, which aims to demonstrate a 0.1 Gbit/s data rate between CubeSats separated by a distance of 1500 km \citep{rodiger2025cubeisliod}. We assume an antenna diameter of 1 m, a receiver noise temperature of 120 K, and an aperture efficiency of 0.7. We simulate SHARPEx observations assuming Medium Earth Orbits (radius 13892 km) as envisioned for SHARP (the simulation results will in principle also hold for LEO, but with a limited maximum baseline length), separated by 30 km. As a ground-truth source model, we take a GRMHD simulation originally intended for M87* at 43 GHz \citep{Davelaar2019}, which we scale to the approximate size and flux of 3C279 as measured by RadioAstron at 22 GHz. While this scaling results in a black hole angular size that is too large, for this test we are interested in SHARPEx’s general ability to recover image structures at these scales. 

With these system and source parameters, the main bottleneck for image quality is the maximum baseline length. In \autoref{fig:sharpsims} below, we keep all system parameters constant while varying the baseline length cutoff between 1500 km \citep[CubeISL demonstration,][]{rodiger2025cubeisliod} and 25,000 km (maximum for these orbits). We simulate synthetic datasets and reconstruct images using \texttt{svlbisim} and the \texttt{eht-imaging} tools \citep{Chael2018}. For the RadioAstron benchmark, we use the observed 3C279 data and simulate observations of our source model with the exact same uv-coverage and thermal noise. 

\begin{figure}[t]
    \centering
    \includegraphics[width=0.24\textwidth]{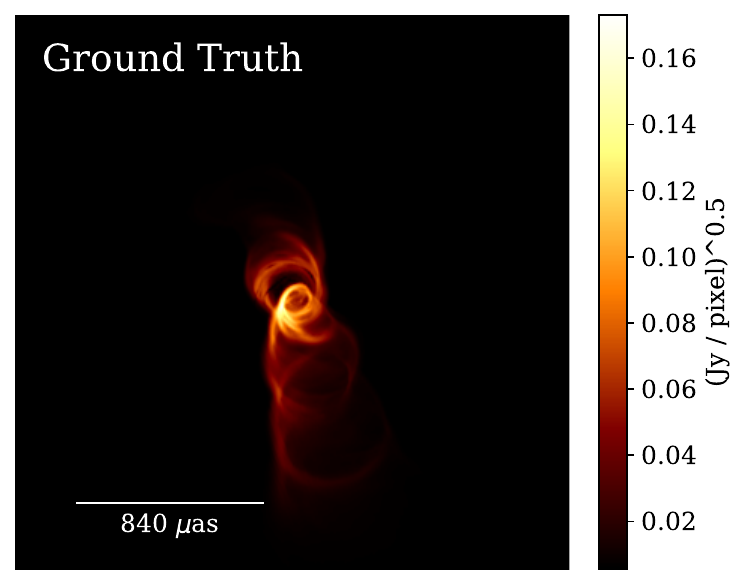}
    \includegraphics[width=0.24\textwidth]{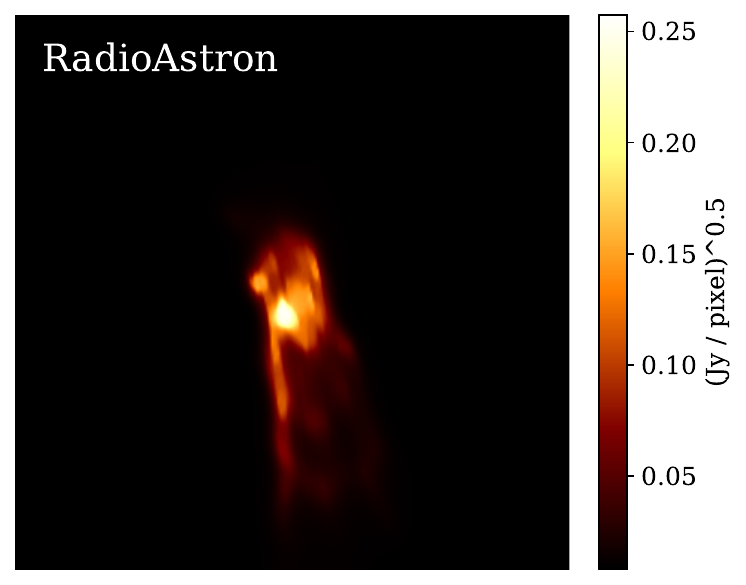}
    \includegraphics[width=0.24\textwidth]{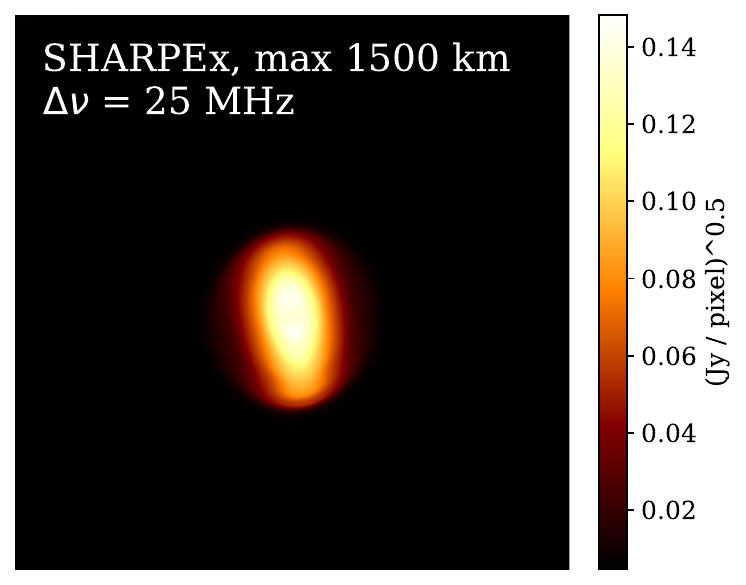}
    \includegraphics[width=0.24\textwidth]{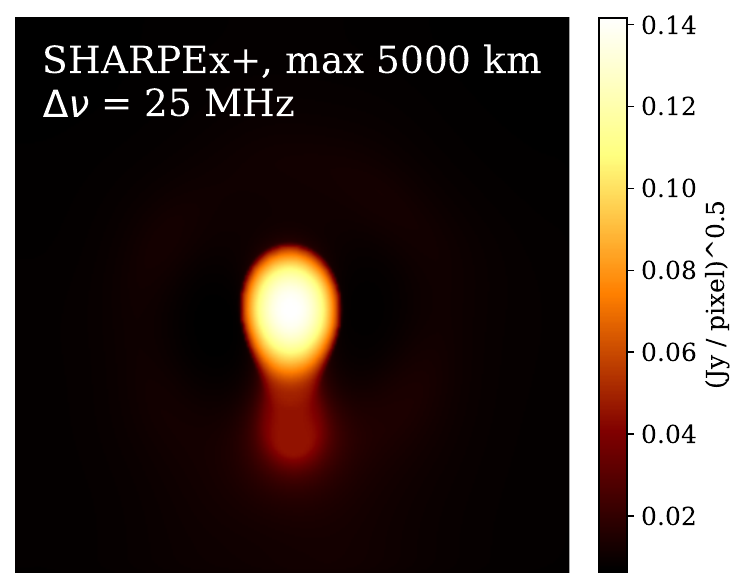}\\
    \includegraphics[width=0.24\textwidth]{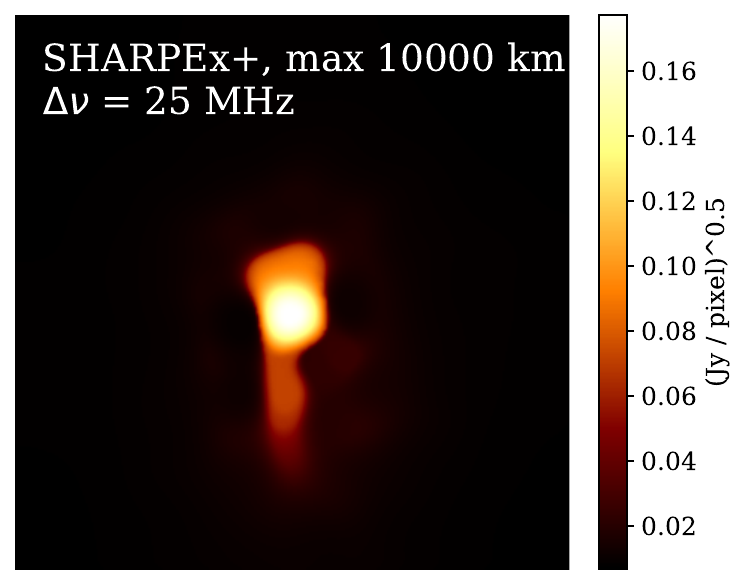}
    \includegraphics[width=0.24\textwidth]{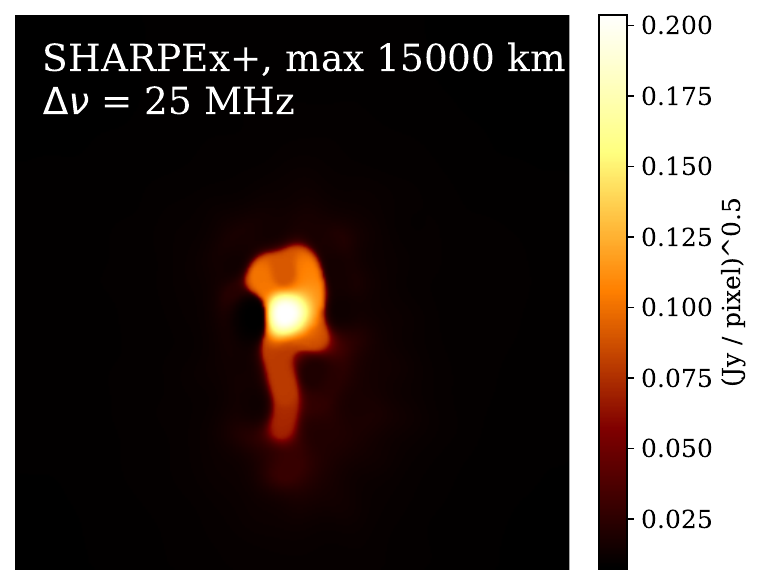}
    \includegraphics[width=0.24\textwidth]{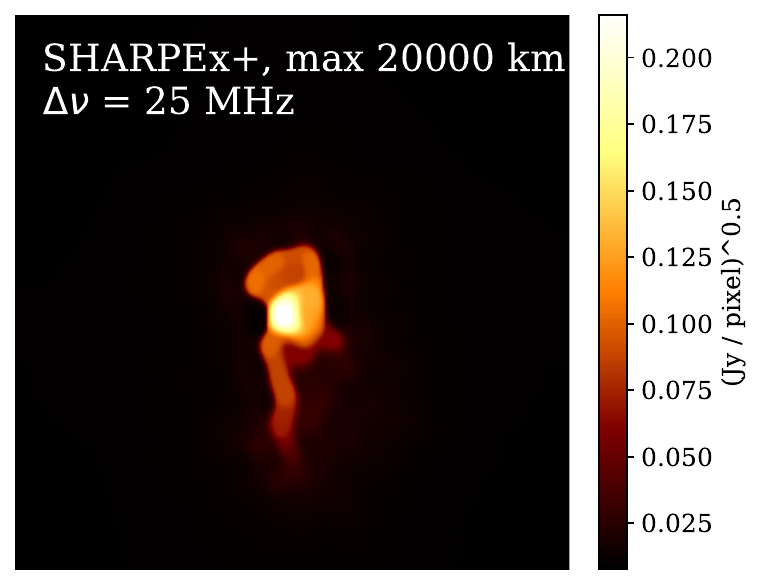}
    \includegraphics[width=0.24\textwidth]{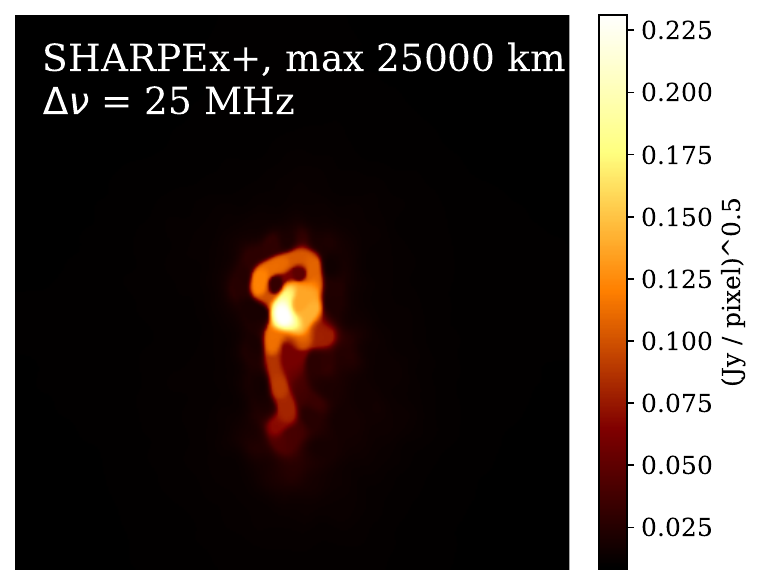}\\
    \includegraphics[width=0.24\textwidth]{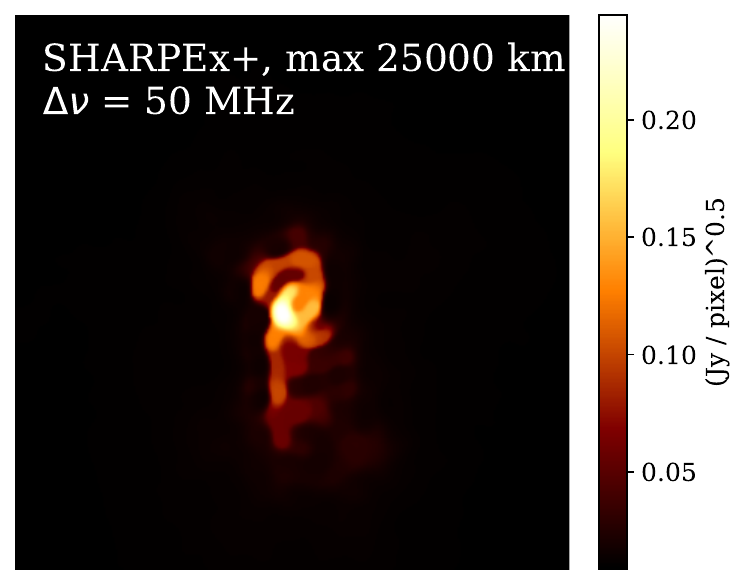}
    \includegraphics[width=0.24\textwidth]{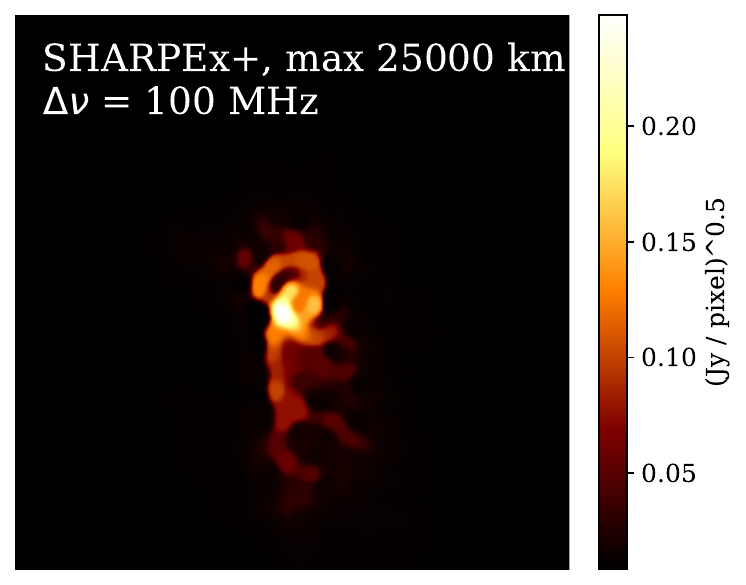}
    \includegraphics[width=0.24\textwidth]{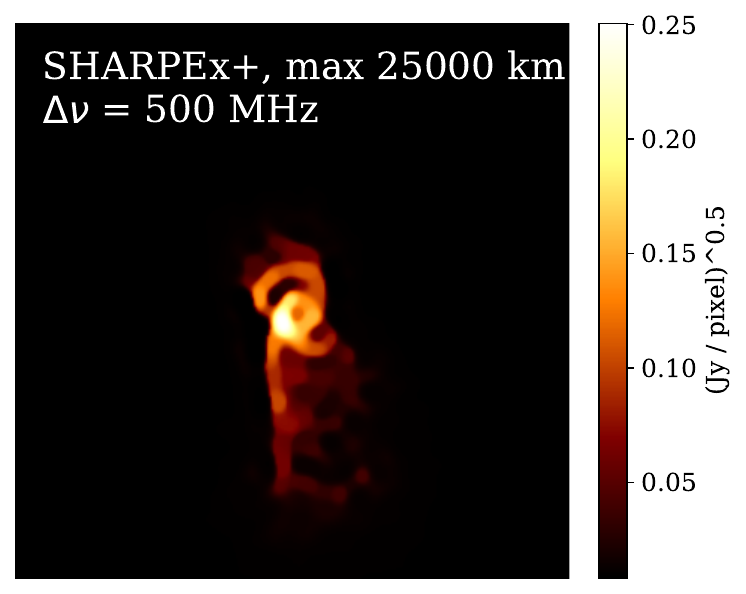}
    \includegraphics[width=0.24\textwidth]{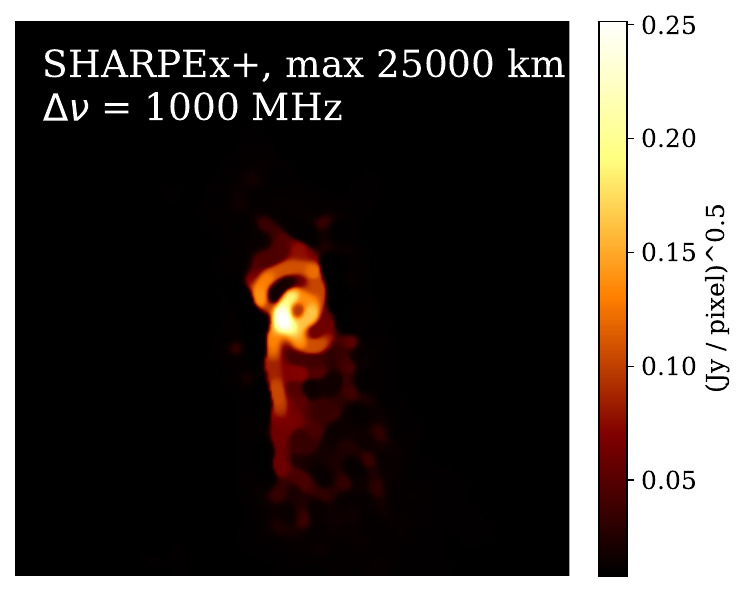}    
    \caption{RadioAstron, SHARPEx, and SHARPEx+ imaging simulations at 22 GHz, for different maximum baseline lengths and frequency bandwidths. While SHARPEx as a low-cost technical demonstrator on CubeSats will most likely be limited to the CubeISL/Cube1G range of 1500 km and 25 MHz bandwidth (top row, third panel), a follow-up mission with enhanced ISL range and bandwidth on larger platforms would enable AGN imaging with unprecedented quality (``SHARPEx+'' panels). The ground-truth image (top left) is an M87* GRMHD frame \citep{Davelaar2019} that has been scaled to match the approximate angular size and flux (27 Jy) of 3C279 \citep{Fuentes2023}. Despite the much smaller SHARPEx dishes (1 m diameter) compared to RadioAstron (10 m diameter) and the much shorter baselines (compare to 100,000+ km for RadioAstron), a SHARPEx-type array may recover features that cannot be reconstructed with RadioAstron's sparse $uv$-coverage, depending on ISL capabilities.}
    \label{fig:ehipimaging}
\end{figure}

The resulting images in \autoref{fig:ehipimaging} show that the maximum distance over which the ISL can carry a large data rate is indeed an important driving factor for the attainable image quality. With 1500 km baselines and 25 MHz bandwidth, detailed structures beyond mas scales cannot be imaged, although the successful recovery of the asymmetric jet structure indicates that these parameters are indeed sufficient to reach our space-to-space VLBI demonstration objectives \autoref{sec:objectives}. 

Pushing the maximum baseline to 15,000+ km and increasing the bandwidth, the image structure recovery is similar to or at the longest baselines even significantly better than the RadioAstron simulation: the ring structures near the black hole, which are invisible in the RadioAstron simulation, become visible in the SHARPEx+ reconstructions at the longest baselines. A bandwidth of 500+ MHz allows high dynamic range imaging of the detailed jet structure further out.

This result is remarkable, because the longest SHARPEx+ baselines (in MEO) are still shorter than those of RadioAstron by an order of magnitude. In addition, SHARPEx+ consists of only two 1-meter dishes, while RadioAstron has a 10-meter dish operating with a full ground array. The reason SHARPEx+ performs so well is its dense and uniform $uv$-coverage (\autoref{fig:uvcov}): traditional ground-based or ground-space VLBI inevitably results in large gaps in the $uv$-plane where no Fourier components are sampled. This problem is especially severe for RadioAstron because of its highly elliptical orbit, so that space-ground baselines are only sampled in a single direction, and most of the $uv$-plane is empty. A space-only array like SHARPEx can sample all Fourier components even with only two antennas. 

With a highly capable ISL, a follow-up mission of SHARPEx will therefore not only demonstrate the space-to-space VLBI technique, but immediately open up a completely new regime for imaging bright AGN, allowing detailed studies of jet launching and collimation.

\section{SUMMARY AND OUTLOOK}
SHARPEx will demonstrate key enabling technologies for the SHARP mission concept to produce razor-sharp images of black holes and directly measure their spacetime metrics to high precision. SHARPEx and SHARP will be the first steps towards a Space Array observatory, with more stations and hence higher sensitivity, opening the door to a breadth of science applications.

SHARPEx will be cost-effective and can be launched on a short timescale due to the use of small (CubeSat or SmallSat) platforms carrying mostly off-the-shelf components. It will image bright AGN at cm wavelengths, and a follow-up science mission even has the potential to provide unprecedented image quality in this frequency regime, allowing detailed studies of the process of jet launching by supermassive black holes. Our ongoing ESA OSIP study will produce a preliminary system design for the SHARPEx demonstrator, which may be followed up by a Concurrent Design Facility (CDF) study at ESA, and form the basis for a mission proposal.

\acknowledgments 
This work is supported by the European Research Council (ERC) Synergy Grant ``BlackHolistic:  Colour Movies of Black Holes: Understanding Black Hole Astrophysics from the Event Horizon to Galactic Scales'' (grant 10107164). This research is supported by the DFG research grant “Jet physics on horizon scales and beyond” (Grant No. 443220636) within the DFG research unit “Relativistic Jets in Active Galaxies” (FOR 5195). Numerical simulations and calculations have been performed on MISTRAL at the Chair of Astronomy at the JMU Wuerzburg.

\bibliography{report} 
\bibliographystyle{spiebib} 
\end{document}